\documentclass[11pt]{article}
\usepackage{geometry}
\usepackage[parfill]{parskip}    % Activate to begin paragraphs with an empty line
\usepackage{graphicx}
\usepackage{amssymb}
\usepackage{epsfig}
\usepackage{epstopdf}
\usepackage{latexsym}
\usepackage{amsmath}
\usepackage{feynmf}
\usepackage{psfrag}
\usepackage{bbm}
\usepackage{cite}
\def\gev{\, \mbox{GeV}}

\def\KL{K\"all\'en-Lehmann}
\def\be{\begin{equation}}
\def\ee{\end{equation}}
\def\bea{\begin{eqnarray}}
\def\eea{\end{eqnarray}}

\begin{document}

\begin{titlepage}
\begin{flushright}
FR-PHENO-2011-006.
\end{flushright}
\vskip 2 cm
\begin{center}

{\Large \bf New spectra in the HEIDI Higgs models}

\vspace{0.5cm}

{\bf J.~J.~van~der~Bij and  B.~Puli\c{c}e}

\vspace{.8cm}

%$^1$
{\it Institut f\"ur Physik, Albert-Ludwigs Universit\"at Freiburg\\
H. Herderstr. 3, 79104 Freiburg i.B., Germany}

\end{center}
\vspace{1cm}
%%%%%%%%%%%%%%%%%%%%%%%%%%%%%%%%%%%%%%%%
\begin{abstract}
\medskip
We study the so-called HEIDI models, which are renormalizable extensions of the standard model
with a higher dimensional scalar singlet field. As an additional parameter we consider a higher-dimensional mixing mass
parameter. This leads to enriched possibilities compared to a previous study. We find effective spectral densities
of the Higgs propagator, consisting of one, two or no particle peaks, together with a continuum.
We compare with the LEP-2 data and determine for which range of the 
model parameters the data can be described. Assuming two peaks to be present we find for the new mass scale
$\nu\approx 56\pm12 \gev$, largely independent of the dimension.
In the limiting case of $d\rightarrow 6$ and two peaks we find a higher
dimensional coupling constant $\alpha_6=0.70 \pm 0.18$, indicative of
strong interactions among the higher dimensional fields. The LHC will not be able to study this Higgs field. 
\end{abstract}

\end{titlepage}

%%%%%%%%%%%%%%%%%%%%%%%%%%%%%%%%%%%%%%%%%
\section{Introduction.}
With the  developments from high energy colliders like LEP
and the Tevatron the standard model (SM) has been established up to
the loop level.  The main missing ingredient is the direct detection of
the SM Higgs boson. The four LEP experiments ALEPH, DELPHI, L3 and OPAL
have extensively searched for the Higgs boson. The final combined result
has been published in \cite{lephiggs}. The absence of a clear signal has led to
a lower limit on the Higgs boson mass of 114.4 GeV at the 95\% confidence level.

%There is a $2.3\,\sigma$ effect seen by all experiments at around 98 GeV.
%A somewhat less significant $1.7\,\sigma$ excess is seen around 115 GeV. Finally
%over the whole range $s^{1/2} > 100\gev$ the confidence level is less than
%expected from background.\\
%
% Within the minimal supersymmetric standard model (MSSM)
%and other extensions [2-5], the excesses at 98 GeV and 115 GeV were interpreted as
%evidence for the presence of two Higgs bosons $H_i$ with couplings reduced
%by a factor $\alpha_i$ to matter
%$g^2_{i} = \alpha_i\, g^2_{SM-Higgs}$. We will call such Higgs bosons fractional Higgses
%in the following. The excess at 98 GeV is well described by a 10\% fractional Higgs.
%More precisely \cite{drees} gives limits $0.056 < \alpha_- < 0.114$ and
%a mass range $95\gev < m_{Higgs} < 101\gev$. The second peak at 115 GeV is then
%interpreted as a second Higgs boson with $\alpha_+ = 0.9$. The first peak
%at 98 GeV is rather convincing. The second one at 115 GeV is  compatible with
%the data, but not really preferred as the
%data at 115 GeV are also compatible with pure background with a similar confidence level.
%The data correspond roughly to background plus one half of a Higgs boson,
% however with a large 
%uncertainty. The MSSM fit is therefore not completely satisfactory and it is natural
%to look for other extensions to describe the data.\\  
Although no clear signal was found the data have some intriguing features,
that can be interpreted as evidence for Higgs bosons beyond the standard
model. 
There are excesses at $98 \gev$ and $115\gev$, that were interpreted in [2-5] as evidence
for the presence of two Higgs bosons. However the fits are somewhat unsatisfactory.
Actually the precision measurements leave only very little
space for extensions to the SM, as these tend to spoil the agreement with experiment
due to a variety of effects, one of the most important of which is
the appearance of flavor-changing neutral currents. For instance supersymmetric 
models have to finely
tune a number of parameters. This leaves only one type of extensions that are
safe, namely the
singlet extensions. Experimentally  right handed
neutrino's appear to exist. Since these are singlets a natural extension of the SM
is the existence of singlet scalars too. These will only have a very limited effect
on radiative corrections, since they appear only in two-loop calculations.
The simplest model giving rise to two Higgs bosons is the Hill model\cite{hill}.
For a mini-review on singlet extensions see \cite{mnmsm}.\\

Of great interest are renormalizable extensions with a higher-dimensional singlet scalar.
In such models the Higgs propagator can change significantly from its standard model
form, in particular if the higher dimensions are non-compact. In this case
the Higgs can become very broad, even though it is light and weakly interacting,
while the spectrum is not anymore a single particle, but at least partly
a continuum. Such a Higgs boson cannot be seen at the LHC, because for a light Higgs boson
one needs a narrow peak to have a signal at the LHC. It is for this reason that we call
these models HEIDI models, for hidi(ng) in high-D(imensions), in analogy to
SUSY for supersymmetry. In a previous paper \cite{dilchbij} these models were compared
with the LEP-2 data and it was shown that a good fit was possible, though in a somewhat narrow range
of parameters. In this paper we extend the study of \cite{dilchbij}, by including an extra parameter,
a mass term mixing the higher dimensional fields among each other. This extra term would be generated by renormalization effects in any case and should therefore be included when comparing with the data.
The analysis is straightforward, but gives a number of extra possibilities, i.e. instead 
of only one peak plus a continuum one finds the possibility of having two peaks or even no peak 
at all in the Higgs propagator.

The paper is organized as follows. In section 2 we describe the model and the Higgs propagator.
In section 3 we describe the possibilities for the \KL\, spectral density. In section 4
we compare with the LEP-2 data. In section 5 we discuss the results.

%%%%%%%%%%%%%%%%%%%%%%%%%%%%%%%%%%%%%%%%%%%%
\section{The model and the Higgs propagator.}
\label{density}

The model consists of the standard model with the Higgs field $\Phi$ and in addition a 
scalar field $H$, moving in $d=4+\gamma$ dimensions.
Normally speaking this would lead to a nonrenormalizable theory. However
since the only interaction is of the form
$H\Phi^{\dagger}\Phi$, which is superrenormalizable in four dimensions,
the theory stays renormalizable. An analysis of the power counting of divergences
shows that one can associate the canonical dimension $1+\gamma/2$ to the
$H$-field. This means that the theory stays renormalizable as long as
$\gamma \le 2$. 
When one assumes, that the extra dimensions are compact, for instance a torus,
with radius $R=L/2 \pi$,
one can simply  expand the $d$-dimensional field
$H(x)$ in terms of Fourier modes:
\begin{equation}
H (x) ={1 \over \sqrt{2}\, L^{\gamma/2}} \sum_{\vec{k}} H_{\vec{k}}(x_\mu)\,
e^{i {2 \pi \over L} \vec{k} \vec{x} },  \;\;\;\;\;\;\;\;\;\;\;\;
  H_{\vec{k}}      =   H_{-\vec{k}}^*~~~        
\end{equation}

Here $x_\mu$ is a four-vector, $\vec{x}$ is $\gamma$-dimensional
and the $\gamma$ components of $\vec{k}$ take only integer values.

In terms of the modes $H_i$ the Lagrangian, that we use, is the following:
\begin{eqnarray}
\label{lagrangian2}
L&=&-\frac{1}{2}D_\mu \Phi^\dagger D_\mu \Phi -\frac{1}{2}\sum (\partial_\mu H_k)^2+\frac{M_0^2}{4}\Phi^\dagger \Phi -\frac{\lambda}{8}(\Phi^\dagger \Phi)^2-\sum \frac{m_k^2}{2} H_k^2\nonumber \\
&-& \frac{g}{2} \Phi^\dagger \Phi \sum H_k-\frac{\zeta}{2}\sum H_i H_j ~~~
\end{eqnarray}
%%%%
Here $m_k^2=m^2+ m_{\gamma}^2\,\vec k^2$, where $\vec k$ is a $\gamma$-dimensional
vector, $m_{\gamma}=2\pi/L$ and $m$ is\newline a $d$-dimensional mass term for the field $H$.

This Lagrangian differs from the one in \cite{dilchbij} through the presence of the
$\sum H_i H_j$ term. This term however would be present in general, while it is 
generated via loop effects in the propagator through the $\Phi^\dagger \Phi \sum H_k$
interactions.
The last two terms can be written as so-called brane-bulk terms of the form:
\begin{equation}
 S=\int d^{4+\gamma}x \prod_{i=1}^{\gamma} \delta(x_{4+i})\left(g_B H(x) 
\Phi^{\dagger}\Phi -\zeta_B H(x) H(x)\right)  
 \end{equation}

It is now straightforward to follow the steps of \cite{dilchbij}.
One expands around the minimum of the potential and inverts the quadratic part
of the Lagrangian, in order to get the  propagator. Introducing the parameter
$M^2=\lambda v^2$, where $v$ is the weak scale,
the quadratic piece takes the form:
%%%% 
\begin{eqnarray}
\label{inversematrix2}
P_{ij}^{-1}(q^2)=
\begin{pmatrix}
q^2+M^2 & -gv & -gv & \cdots & -gv \\
-gv & q^2+m_1^2+\zeta & \zeta & \cdots &  \zeta \\
-gv & \zeta & q^2+m_2^2 +\zeta&  & \zeta \\
\vdots & \vdots & & \ddots & \vdots \\
-gv & \zeta & \zeta & \cdots & q^2+m_n^2 +\zeta
\end{pmatrix}~~~
\end{eqnarray}
%%%%
We need to find the Higgs-field propagator which will be the $(1,1)$ component of the matrix $P_{ij}$. For this reason, we use the following determinant formulas:
%%%% 
\begin{eqnarray}
\label{det1}
\begin{vmatrix}
a_1 +\alpha & \alpha & \alpha & \cdots & \alpha \\
\alpha & a_2 +\alpha & \alpha & \cdots & \alpha \\
\alpha & \alpha & a_3 + \alpha &  & \alpha \\
\vdots & \vdots &  & \ddots & \vdots \\
\alpha & \alpha & \alpha & \cdots & a_n + \alpha
\end{vmatrix}=\prod a_n \Bigg(1+\alpha \sum \frac{1}{a_n} \Bigg)~~~
\end{eqnarray}
and
%%%%
\begin{eqnarray}
\label{det3}
\begin{vmatrix}
A & \beta & \beta & \cdots & \beta \\
\beta & a_1 + \alpha & \alpha & \cdots & \alpha \\
\beta & \alpha & a_2 + \alpha & & \alpha \\
\vdots & \vdots & & \ddots & \vdots \\
\beta & \alpha & \alpha & \cdots & a_n +\alpha
\end{vmatrix}= A \prod a_n \Bigg(1-\frac{\beta^2}{A} \sum \frac{1}{a_n} + \alpha \sum \frac{1}{a_n} \Bigg)~~~
\end{eqnarray}
where the symbols correspond to the following expressions
%%%%
\begin{eqnarray}
A&=&q^2 + M^2 ~~~\nonumber \\
\beta &=&-g v ~~~ \nonumber \\
a_n&=&q^2+m_n^2~~~ \nonumber \\
\alpha&=&\zeta~~~
\end{eqnarray}
%%%%
The ratio of \eqref{det1} to \eqref{det3} gives us the $(1,1)$ component of the inverse of the matrix \eqref{inversematrix2}, \textit{i.e.} the Higgs-field propagator
\begin{eqnarray}
\label{higgsprop3}
D_{HH}(q^2)=\frac{1+\zeta \frac{\Gamma (1-\gamma /2)}{(4 \pi)^{\gamma /2}}(q^2+m^2)^{\frac{\gamma-2}{2}}}
{(q^2 + M^2)\Bigg(1-\frac{\Gamma (1-\gamma /2)}{(4 \pi)^{\gamma/2}}(q^2+m^2)^{\frac{\gamma -2}{2}}
\Bigg(\frac{g^2 v^2}{(q^2+M^2)}-\zeta \Bigg) \Bigg)}
\end{eqnarray}
%%%%
where we have taken the continuum limit
%%%%
\begin{eqnarray}
\sum \frac{1}{a_n}&=&\frac{\Gamma(1-\gamma /2)}{(4 \pi)^{\gamma /2}}(q^2 + m^2)^{\frac{\gamma -2}{2}}
\end{eqnarray} 
%%%%

We make the following definitions:
\begin{eqnarray}
\mu^{8-d}&= g^2 v^2 \frac{\Gamma (1-\gamma /2)}{(4\pi)^{\gamma /2}} \nonumber \\
\nu^{6-d}&=\mid\zeta\mid \frac{\Gamma (1-\gamma /2)}{(4\pi)^{\gamma /2}}
\end{eqnarray}
%%%%
The Higgs-field propagator \eqref{higgsprop3} becomes
%%%%
\begin{eqnarray}
\label{higgsprop4}
D_{HH}(q^2)=\Bigg(q^2 +M^2-\frac{\mu^{8-d}}{(q^2+m^2)^{\frac{6-d}{2}} \pm \nu^{6-d}} \Bigg)^{-1}
\end{eqnarray}
where the sign in front
of the $\nu$ term is the sign of $\zeta$.
 In the rest of the paper $\nu$ is supposed to be a positive number.
In the following we will use the auxiliary quantity $\delta=(6-d)/2$.

\section{The \KL\, spectral density.}
Given the form of the Higgs-propagator found in the previous section one can derive
from it the \KL\,\cite{kallen,lehmann} spectral density. The spectral density $\rho(s)$ is simply given by the imaginary part of the
propagator.
%%%%
\begin{eqnarray}
\label{spdensitycondef}
\rho_{\textrm{KL}}(s)=-\frac{1}{2\pi i}\Bigg(D(s+i\epsilon)-D(s-i\epsilon) \Bigg)~~~
\end{eqnarray}
%%%%
In case there is a zero in the inverse propagator one has a pole, that shows up as a
delta-function in the \KL\, density. This is the case one is used to in ordinary field theory
where the pole corresponds to a particle on the mass-shell. In the HEIDI models
the spectrum is more general containing poles and a continuum. Different possibilities
arise for different values of the parameters $m,M,\mu,\nu$ from \eqref{higgsprop4}.
 However not all possibilities
correspond to a physically allowed spectral density. One must demand that there  are no tachyon
poles in the propagator. Such a pole corresponds to an instability in the theory, that arises
when one does not expand around a minimum of the potential. Typically the potential would not be
bounded from below. Alternatively the attractive force among the Higgs particles due to the
high-D particles is larger than the repulsion from the selfcoupling\cite{dilchbij}. 
Different possibilities are discussed 
in the subsections below. The range $4<d< 6$ can be treated as a whole.
The limiting case $d\rightarrow 6$ is treated separately.
The continuum part $\rho_c(s)$ of the spectral density for $4<d<6$ is given by
\begin{equation}
\rho_{c}(s) = \frac{1}{\pi} \frac{\mu^{8-d}\, (s-m^2)^\delta\, \sin(\pi\delta)\,\, \theta(s-m^2)}
{
\left\{(M^2-s)[(s-m^2)^\delta\cos(\pi\delta)\pm\nu^{2\delta}]-\mu^{8-d}\right\}^2
+ \sin^2(\pi\delta)(M^2-s)^2(s-m^2)^{2\delta}
}
\end{equation}

\subsection{Negative sign in front of the \boldmath$\nu$ term.}
In the case the $\nu$ term appears with a negative sign,
 the condition for the absence of a tachyon pole
is:\\

No tachyon: $M^2 (m^{6-d} -\nu^{6-d}) > \mu^{8-d}$.

Assuming the tachyon condition is fulfilled we consider several cases.

$M^2>m^2$ : there is one pole in the propagator.\\

$M^2<m^2$ : there is one pole in the propagator if:
$\nu^{6-d}\,(m^2-M^2)\, <  \mu^{8-d}$.\\
$M^2<m^2$ : there are two poles in the propagator if:
$\nu^{6-d}\,(m^2-M^2)\, > \mu^{8-d}$.\\

\subsection{The case \boldmath$\nu=0$.}
The case $\nu=0$ was dicussed in \cite{dilchbij}.\\
The no-tachyon condition reads:
$M^2 m^{6-d} > \mu^{8-d}$.\\
In this case there is always one pole plus a continuum.

\subsection{Positive sign in front of the \boldmath$\nu$ term.}
In the case of a positive sign in front of the  $\nu$ term the conditions are the following:

No tachyon: $M^2 (m^{6-d} + \nu^{6-d}) > \mu^{8-d}$.\\

$M^2>m^2$:  no pole if: $\nu^{6-d}\,(M^2-m^2)\,> \mu^{8-d}   $.\\
$M^2>m^2$: one pole if: $\nu^{6-d}\,(M^2-m^2)\,< \mu^{8-d}   $.\\

$M^2<m^2$: always one pole.\\

\subsubsection{\boldmath$m=0$.}
In the theory we have four mass parameters $m,M,\mu,\nu$.
This makes for a fairly complicated analysis and one might wonder
whether one could eliminate one or more of the parameters, without getting
into trouble with renormalizability. Of course one could put $\mu=0$, but
this simply decouples the higher dimensions from the four dimensional fields
and one  gets the standard model back. More interesting is to put
$m=0$. This is actually consistent with renormalization, as the higher dimensional
mass term does not get generated through the interactions with the ordinary Higgs field.
Only the $\nu$-term is genererated by loop effects.  In the case that $\nu>0$
one can indeed put $m=0$, without getting a tachyon in the propagator.
The  
no tachyon condition becomes: $M^2\, \nu^{6-d}>\mu^{8-d}$.
This implies that there is no pole in the propagator and one has a continuum only.
The situation is somewhat similar to the ideas in \cite{georgi}, where the singlet
fields are to be conformal invariant, however the conformal invariance is here
broken by the $\mu,\nu$ and $M$ terms. This is necessary in order to get a satisfactory 
spectrum\cite{heidiun}.

In discussions the question often arises, whether putting a parameter like $m$
equal to zero, should be considered an unnatural fine-tuning, since
this relation is not due to a symmetry.
It must be emphasized, that the condition is preserved under renormalization.
Therefore one cannot say that imposing $m=0$ is really arbitrary,
 as long as one has no deeper
insight in some form of underlying dynamics. The question is somewhat philosophical:
should one only consider the terms that are\,\, {\it needed for}\,\, renormalizability or assume
all terms to be present that are\,\, {\it allowed by}\,\, renormalizability? 
In most ordinary models these classes are the same, so the question does not arise.

\subsection{The case \boldmath$d\rightarrow 6$.}
For the critical dimension $d=6$ one cannot simply copy the formulas
from section 2. In $d=6$ the $\Gamma$-function develops a pole, so its value
cannot be absorbed in the definition of the mass parameters, that end up in the 
Higgs propagator. Instead one has to use a limiting procedure. Another way to see the 
problem is by noting that the parameter $\nu$ becomes dimensionles and is therefore 
to be considered as a coupling constant and not  as a mass parameter. It was noted before
\cite{wise,dilcher,aguila,derham} that
in $d=6$ the $\nu$ parameter has a renormalization group running already at the tree level.
There are different ways to write the propagator in this case, depending on how one defines
the new parameters in the propagator.
We choose a new variable $\alpha_6$ to describe the propagator. The name is chosen to
make clear that we are talking about a coupling constant. In the formulas $\alpha_6$ can be
taken to be positive, negative or zero, zero corresponding to the case without $\nu$ term. 
The form of the Higgs propagator becomes

\begin{eqnarray}
\label{higgsprop2}
D_{HH}(q^2)=\Bigg(q^2+M^2 + \mu^{2} \frac{\log((q^2+m^2)/m^2)}{1 + \alpha_6\,\log((q^2+m^2)/m^2)}\Bigg)^{-1}
\end{eqnarray}

In this form $\alpha_6=0$ reproduces the propagator from \cite{dilchbij}.
The analysis is simple. The absence of a tachyon requires
$$\alpha_6 \geq 0 \hskip 0.3cm {\rm and}\hskip 0.3 cm M^2 \geq 0$$

Furthermore one has

$\alpha_6\,(m^2 - M^2) <   \mu^2$ : there is one pole in the propagator.\\
$\alpha_6\,(m^2 - M^2) \geq \mu^2$ : there are two poles  in the propagator.\\

Because of the possibility of having two poles the limit  $d\rightarrow 6$ 
is similar to the $\nu < 0$ case in the range $4<d<6$.
Putting $m^2=0$ in the $q^2+m^2$ term would lead to a tachyon in the propagator.\\
The continuum spectral density $\rho_c(s)$ is here given by:
\begin{equation}
\rho_c(s) = \frac{\mu^2\,\,\theta(s-m^2) }
{\left[-s+M^2+(\mu^2-\alpha_6 s+\alpha_6 M^2)\log((m^2-s)/m^2)\right]^2 + 
\pi^2\,(\mu^2-\alpha_6 s+\alpha_6 M^2)^2}
\end{equation}

%\subsection{$\nu=0$}

%%%%%%%%%%%%%%%%%%%%%%%%%%%%%%%%%%%%%%%%%%%%%%%%%%%%%%%%%%%%%%%%%%%%%%%%%%%%%%%%%%%%%%%%%%%%%%%%%%%%%%%%%%%%%%%%%%
%%%%%%%%%%%%%%%%%%%%%%%%%%%%%%%%%%%%%%%%%%%%%%%%%%%%%%%%%%%%%%%%%%%%%%%%%%%%%%%%%%%%%%%%%%%%%%%%%%%%%%%%%%%%%%%%%%
%\newpage

\section{Comparison with the LEP-2 data}
\subsection{Description of the data}
As mentioned in the introduction the LEP-2 Higgs search data 
have some features, that make them interpretable within our model.
 There is a $2.3\,\sigma$ effect seen by all experiments at around 98 GeV.
A somewhat less significant $1.7\,\sigma$ excess is seen around 115 GeV. Finally
over the whole range $\sqrt{s} > 100\gev$ the confidence level is less than
expected from background. Within the minimal supersymmetric standard model (MSSM)
and other extensions [2-5], the excesses at 98 GeV and 115 GeV were interpreted as
evidence for the presence of two Higgs bosons $H_i$ with couplings  to matter reduced
by a factor $\alpha_i$, giving
$g^2_{i} = \alpha_i\, g^2_{SM-Higgs}$. We will call such Higgs bosons fractional Higgses
in the following. The excess at 98 GeV is well described by a 10\% fractional Higgs.
More precisely \cite{drees} gives limits $0.056 < \alpha_- < 0.114$ and
a mass range $95\gev < m_{Higgs} < 101\gev$. The second peak at 115 GeV is then
interpreted as a second Higgs boson with $\alpha_+ = 0.9$. The first peak
at 98 GeV is rather convincing. The second one at 115 GeV is  compatible with
the data, but not really preferred as the
data at 115 GeV are also compatible with pure background with a similar confidence level.
Actually the data correspond roughly to background plus one half of a Higgs boson,
 however with a large uncertainty.\\

Within our model there are different possibilities for the spectrum.
One can have one, two or no peaks plus a continuum in the spectrum.
We propose different ways to analyze the data for the different cases.
In the case of two peaks we have, for a fixed dimension $d$, four
parameters $M,m,\mu,\nu$ and can use the strength and location of the 
two peaks to completely fix these parameters. This can be straightforwardly done.
Our model has the advantage over the MSSM, that the sum of the strenghts of the peaks
does not have to be one, allowing for more freedom.
Hereby we ignore the possibility of a 
continuum between the peaks, which is not very pronounced in the data.

Another way to analyze the data was presented in\cite{dilchbij}.
If there is one peak plus a continuum one can do the following.
One takes the excess at 98 GeV at face value and interprets it as the delta-peak
in the propagator. The excess at 115 GeV is interpreted as an enhancement
due to the continuum of the Higgs propagator. Because of the uncertainty of this
excess we will only demand that a Higgs integrated spectral
density $\int \rho(s) ds > 30\%$
is present in the range $110\gev < \sqrt{s} < 120\gev$. 

In the case of no peaks one could take the special case $m=0$, allowing for a spectrum
extending down to $m=0$. One could then worry, whether a possible continuum
Higgs propagator has simply been overlooked, because the density is too low for 
the sensitivity of the LEP-2 experiments. One might hope to get a better fit
to the LEP-1 indirect determination of the Higgs mass, which is quite low, about 89\gev.
However the published data have not been analyzed with this point of view in mind.
There are limits in the literature on the strength of a single Higgs
at every mass \cite{lephiggs}. However these cannot be used to analyze our model,
as the size of the different mass bins is not given, so there is no way to scale the
mass density. A precise comparison would need a complete reanalysis of the data.
So we will not pursue this possibility here any further.
Also the analyses with one or two peaks suffer from the fact that the data are not very precise 
and have not been analyzed with this type of model in mind.
In particular the statistical significance of the deviations of the standard model
is hard to determine. Rough estimates give about $3.0-3.4\, \sigma$.

\subsection{Fit to the two-pole models.}
\subsubsection{Parameters and data.}
When one tries to fit the parameters to the LEP-2 data,
one finds that both $M$ and $m$ are quite close to the
location of the highest peak. It is therefore
useful to define some auxiliary quantities.

We define:\begin{eqnarray}
m^2 &=& s_+ + \epsilon (s_+ - s_-) \nonumber \\
M^2 &=& s_+ - \Delta (s_+-s_-)
\end{eqnarray}

Here $s_+$ is the location of the most massive peak, with
a contribution $\alpha_+$ in the spectral density; 
$s_-$ is the location of the less massive peak, with
a contribution $\alpha_-$ in the spectral density.

For a fixed dimension we have four data points to fit four parameters,
 that are therefore completely fixed.
We vary the input values over the following range:

\begin{eqnarray}
0.056 &< \alpha_- &< 0.144 \nonumber \\
0.3 &< \alpha_+ &< 0.7 \nonumber \\
95\gev &< \sqrt{s_-} &< 101\gev \nonumber \\
111\gev &< \sqrt{s_+} &< 117\gev
\end{eqnarray}

\subsubsection{Solution in the range $4<d<6$.}

One first finds $\epsilon$ from the equation:
\begin{equation}
\label{epseq}
\delta^2 \cdot \frac{\alpha_-}{1-\alpha_-} \cdot \frac{\alpha_+}{1-\alpha_+} =
\epsilon(1+\epsilon)\, 
\left[\left(\frac{1+\epsilon}{\epsilon}\right)^{\delta/2}
-\left(\frac{\epsilon}{1+\epsilon}\right)^{\delta/2}
\right]^2
\end{equation}
For finite $\delta$,  away from $d=6$ and small $\epsilon$\\
a good approximation $\epsilon_{app}$ to the solution is:
\begin{equation}
\epsilon_{app} =\left( \delta^2 \cdot \frac{\alpha_-}{1-\alpha_-} \cdot \frac{\alpha_+}{1-\alpha_+}\right)
^{\frac{2}{d-4}}
\end{equation}

Next one can determine $\Delta$ from the equation:

\begin{equation}
\frac{\alpha_+}{1-\alpha_+} \cdot \frac{1-\alpha_-}{\alpha_-} . 
\left(\frac{1+\epsilon}{\epsilon}\right)^{\frac{d-4}{2}} =
(\frac{1-\Delta}{\Delta})^2 
\end{equation}

Next one has :
\begin{equation}
\mu^{8-d} = \Delta(1-\Delta) (s_+-s_-)^{1+\delta}\left((1+\epsilon)^{\delta}
-\epsilon^{\delta}\right) 
\end{equation}
and finally:
\begin{equation}
\nu^{6-d} = (m^2-s_-)^{\delta}-\frac{\mu^{8-d}}{M^2-s_-}= 
(m^2-s_+)^{\delta}-\frac{\mu^{8-d}}{M^2-s_+} 
\end{equation}

Trying to fit to the LEP-2 data we have  small values 
for $\alpha_-$ and $(s_+-s_-)/s_+$. One can take
$\epsilon \approx 0$ and find an approximate solution:
\begin{eqnarray}
\Delta&=&\frac{\delta.\alpha_-}{1-\alpha_-+\delta.\alpha_-}\\
\mu&=&[\Delta(1-\Delta)]^{\frac{1}{8-d}}\sqrt{s_+-s_-}\\
\nu&=&[1-\Delta]^{\frac{1}{6-d}}\sqrt{s_+-s_-}\\
\end{eqnarray}
As also $\Delta$ is small the scale of $\nu$ is basically fixed
independent of the dimension to about:
\begin{equation}
\nu = 56 \pm 12 \gev
\end{equation}

This approximation works very well as long as one is not close
to the limit $d\rightarrow6$. In the case $d=4$ one reverts to the
original Hill model, that has two peaks and no continuum. As in this case there
is only one singlet mode, the presence or absence of the $\nu$ term
makes no difference. 

\subsubsection{Solution for the limit $d\rightarrow 6$.}

In the limiting case $d\rightarrow 6$ the term on the right side
of equation \ref{epseq} behaves like $\epsilon\log^2(\epsilon)$. 
In this case one should not use the approximation
$\epsilon \rightarrow 0$, but use the exact equations.

One first finds $\epsilon$ from the equation:
\begin{equation}
\frac{\alpha_-}{1-\alpha_-} \cdot \frac{\alpha_+}{1-\alpha_+} =
\epsilon(1+\epsilon)\, \log^2( \frac{\epsilon}{1+\epsilon})
\end{equation}
Next one can determine $\Delta$ from the equation:

\begin{equation}
\frac{\alpha_+}{1-\alpha_+} \cdot \frac{1-\alpha_-}{\alpha_-} \cdot \frac{1+\epsilon}{\epsilon} =
(\frac{1-\Delta}{\Delta})^2 \left( \frac{\log(\epsilon) + \log((s_+-s_-)/m^2)}
{\log(1+\epsilon) + \log((s_+-s_-)/m^2)}                   
\right)^2
\end{equation}

Next one has :
\begin{equation}
\mu^2 = \Delta(1-\Delta) (s_+-s_-) \left(\frac{1}{\log((m^2-s_+)/m^2)}-\frac{1}{\log((m^2-s_-)/m^2)}\right)
\end{equation}
and finally:
\begin{equation}
\alpha_6 = -\frac{\mu^2}{M^2-s_-} - \frac{1}{\log((m^2-s_-)/m^2)} = 
-\frac{\mu^2}{M^2-s_+} - \frac{1}{\log((m^2-s_+)/m^2)}
\end{equation}

We use these formulas to determine the values of $m,M,\mu,\alpha_6$ from the location and the strengths
of the peaks. 
The values of $m$ and $M$ are always close to $\sqrt{s_+}$. For the range 
of $\mu$ we find $6\gev < \mu < 22\gev$. For $\alpha_6$ we find
$0.52 < \alpha_6 < 0.88$.
In particular the last number is interesting. It would be the first measurement
of a higher-dimensional coupling constant. The value indicates a rather strong coupling.

\subsection{Models with one pole and a continuum.}
\subsubsection{Fitting the data.}
In this case we follow the procedure from \cite{dilchbij}.
We will take the excess at 98 GeV at face value and interpret it as the delta-peak
in the propagator. The excess at 115 GeV is interpreted as an enhancement
due to the continuum of the Higgs propagator. Because of the uncertainty of this
excess we will only demand that a Higgs integrated spectral
density $\int \rho(s) ds > 30\%$
is present in the range $110\gev < s^{1/2} < 120\gev$. 
The delta-peak will be assumed to correspond to the
peak at 98 GeV, with a fixed value of $\alpha_-$.
Ultimately we will vary the location of the peak between
$95\gev < \sqrt{s_-} < 101 \gev$ and the strength of the peak between
 $0.056 < \alpha_- < 0.144$.
After fixing $\alpha_-$ and $m_-$ we have two free variables,
 which we take to be $\mu$ and $\nu$. If we also assume a value for
 $\mu$  and for $\nu$ all parameters and thereby the
spectral density is known. We can then numerically integrate the
spectral density over selected ranges of $s$. The allowed range of $\mu$ 
and $\nu$
is subsequently determined by the data at 115 GeV.
Since the peak at 115 GeV is not very well constrained, we
demand here only that the integrated spectral density
from $s_{down} = (110\gev)^2$ to $s_{up} = (120\gev)^2$
is larger than 30\%.  In general the continuum starts
very close to the lowest peak.
This allows for a natural explanation, why the CL for the fit in the
whole range from 100 GeV to 110 GeV is somewhat less than what is expected by
pure background. The enhancement can be due to a slight, spread-out Higgs signal.
Actually when fitting the data with the above conditions sometimes
 the integrated spectral density in the range
100 GeV to 110 GeV can become rather large, which would lead to problems
with the 95\% CL limits in this range. We therefore additionally demand
that the integrated spectral density in this range is less than 30\%.
There is no problem fitting the data with these conditions. 
As a final consistency check we have checked that the results are in 
agreement with the upper limit on the Higgs boson mass from 
precision measurements $m_H~<~158\gev$. The integrated spectral density
above $158 \gev$  is less than a few percent.

\subsubsection{Allowed range in the $\mu,\nu$ or $\mu,\alpha_6$ plane.}

Varying the input parameters as mentioned above, one determines for each dimension an
allowed range in the $M,m,\mu,\nu$ hypervolume.
Given the fact that one has four mass  parameters, 
one could make six different two-dimensional projections.
However they are not equally interesting. The value of $m$ is always close to the location of the peak
at $98\gev$, $M$ is in the neighbourhood of $115\gev$, but can move up 
to about $140\gev$. We therefore only give graphs of the projection
in the $\mu,\nu$-plane which shows the most interesting behaviour.

The allowed range for the parameters $\mu$ and  $\nu$ are given in the figures 1-3
for the dimensions $d=4.5, 5.0, 5.5$.
 In the figures three curves are drawn.
The points  within the outer curve are the points for which there is a solution
within the range $0.056<\alpha_-<0.144$ and $95\gev < \sqrt{s_-} < 101 \gev$.
For the points within the middle curve
there is  a solution with $\alpha_-=0.1$ and $\sqrt{s_-}=98\gev$.
The points inside the inner curve have a solution for all values in the
range $0.056<\alpha_-<0.144$ and $95\gev < \sqrt{s_-} < 101 \gev$. 

For the case $d\rightarrow 6$ the procedure is the same, only the parameters are
now $M,m,\mu,\alpha_6$. We give the projection on the $\mu,\alpha_6$ plane in figure 4.

%\newpage
\begin{figure}
\psfrag{NU}{$\nu$}
\psfrag{MU}{$\mu$}
\psfrag{d=4.5}{{\large d=4.5}}
\centering
\includegraphics[width=0.5\textwidth,angle=-90]{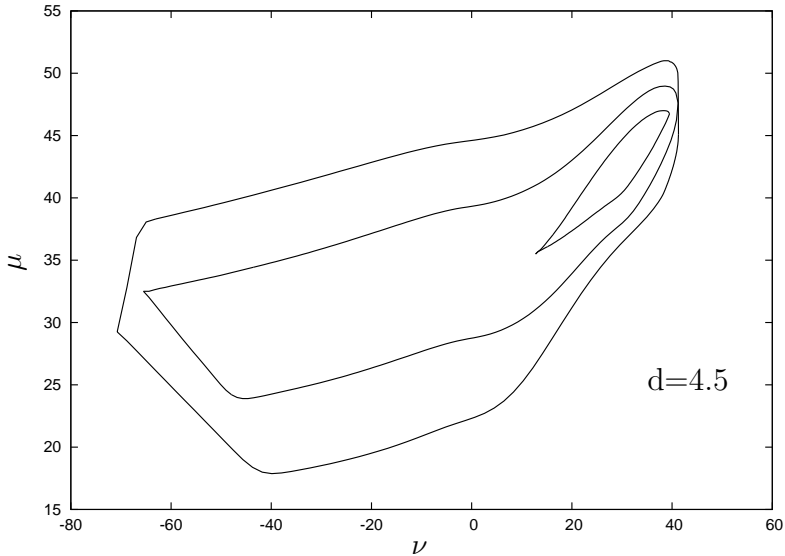}
\caption{\label{fig:01} The allowed $\mu,\nu$-plane for the 4.5 dimensional HEIDI model.}
\end{figure}

\begin{figure}
\psfrag{NU}{$\nu$}
\psfrag{MU}{$\mu$}
\psfrag{d=5.0}{{\large d=5.0}}
\centering
\includegraphics[width=0.5\textwidth,angle=-90]{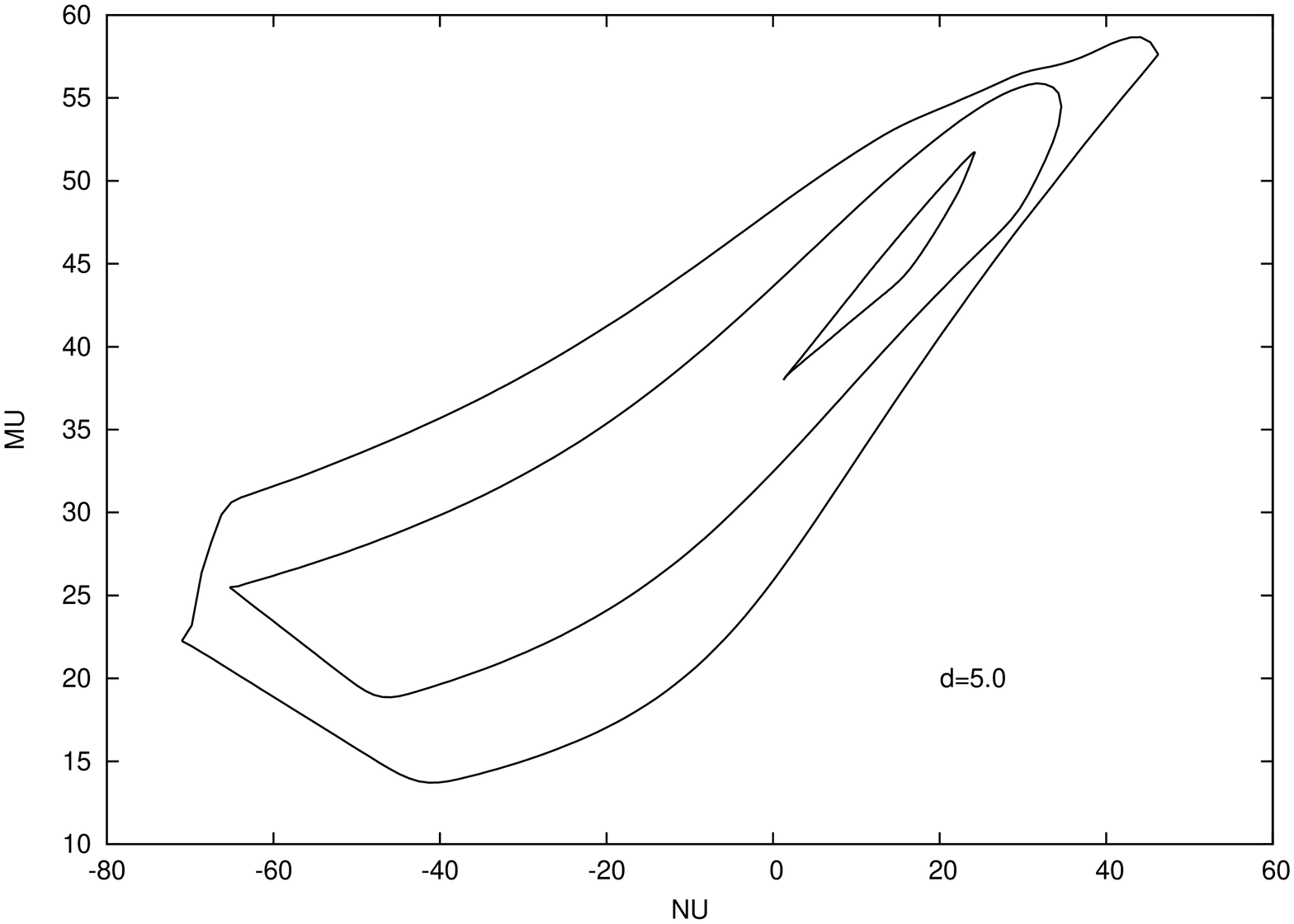}
\caption{\label{fig:02} The allowed $\mu,\nu$-plane for the 5.0 dimensional HEIDI model.}
\end{figure}

\begin{figure}
\psfrag{NU}{$\nu$}
\psfrag{MU}{$\mu$}
\psfrag{d=5.5}{{\large d=5.5}}
\centering
\includegraphics[width=0.5\textwidth,angle=-90]{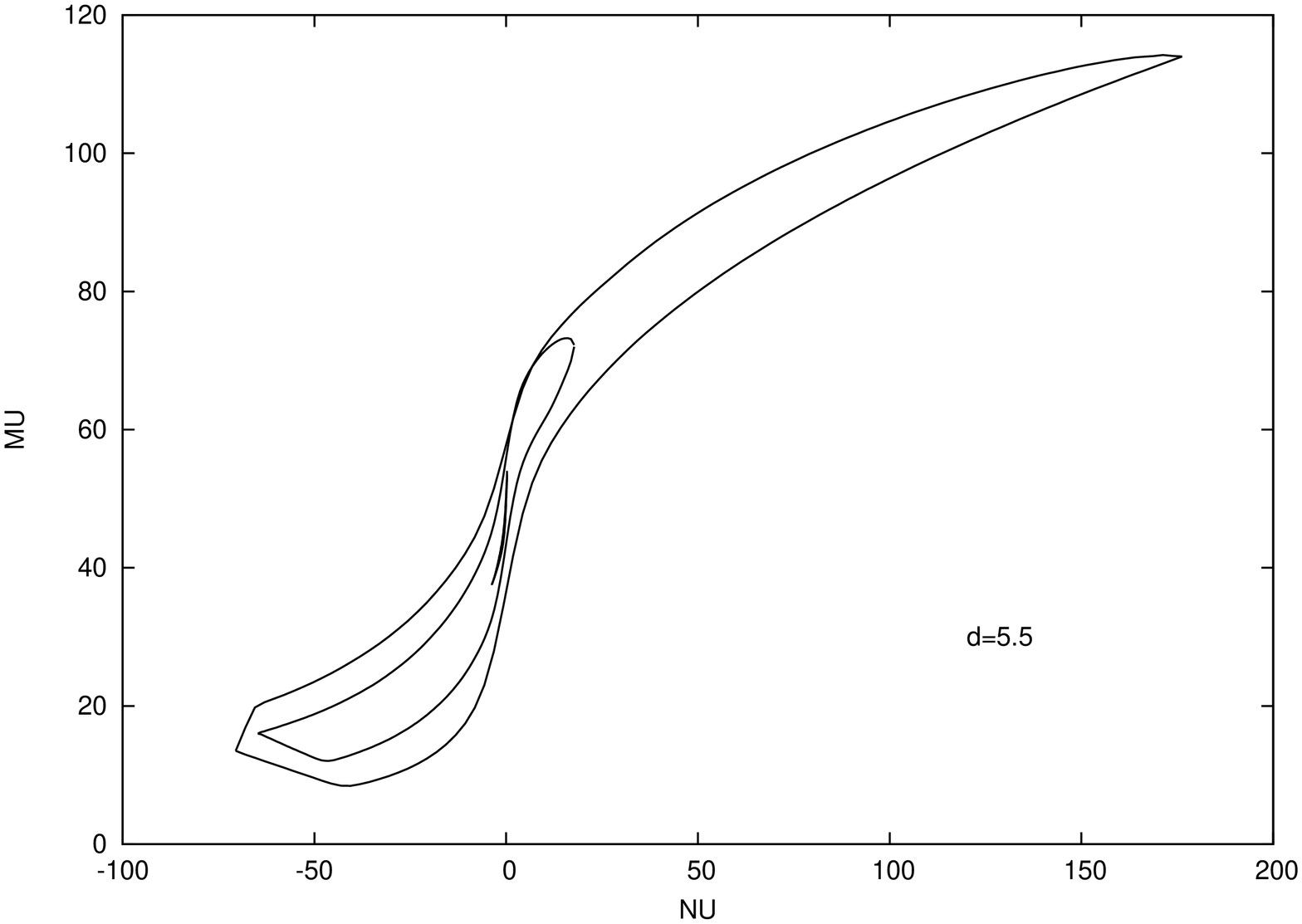}
\caption{\label{fig:03} The allowed $\mu,\nu$-plane for the 5.5 dimensional HEIDI model.}
\end{figure}

\begin{figure}
\psfrag{alpha6}{$\alpha_6$}
\psfrag{MU}{$\mu$}
\psfrag{D=6}{{\large $d\rightarrow 6$}}
\centering
\includegraphics[width=0.5\textwidth,angle=-90]{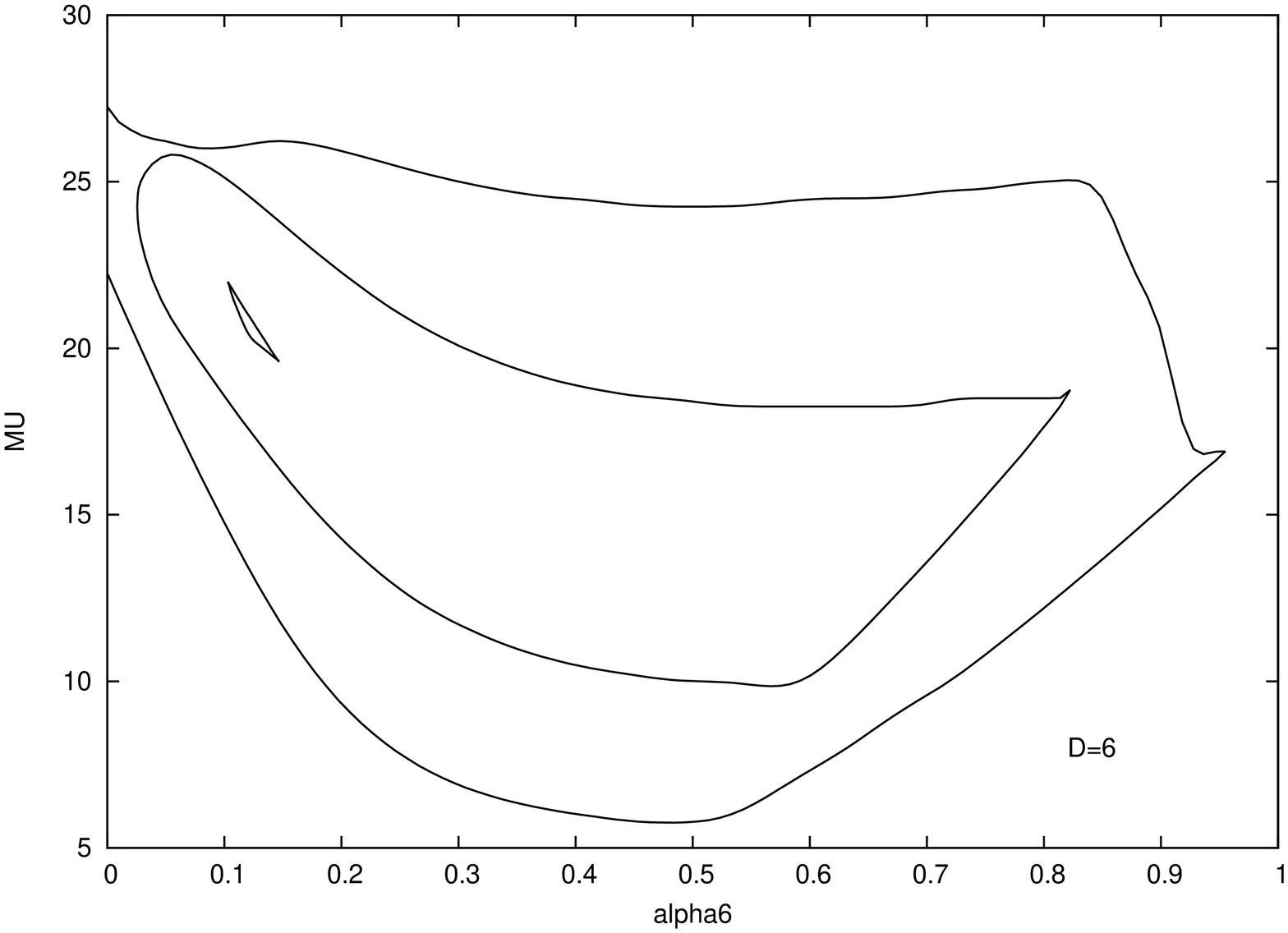}
\caption{\label{fig:04} The allowed $\mu,\alpha_6$-plane for the 6 dimensional HEIDI model.}
\end{figure}

\section{Discussion}

We introduced an extra parameter $\nu$, a mass-mixing term for the
higher dimensional fields,  in the class of HEIDI models
and have shown that this gives rise to a broader range of possibilities
than was studied before.
As a first conclusion, due to the extra parameter,
 we see that the spectra containing a single pole
 can describe the LEP-2 data in a rather broad range of parameters,
without any great form of fine-tuning, so these models are quite natural.
One could even go as far as to say, that this should be the generic form
of electroweak symmetry breaking.
If the data are interpreted as having two peaks, it is interesting, that
the new parameter can be determined rather precisely, largely independent
of the dimension. Unfortunately the data are not very precise, otherwise one
could have claimed the clear  measurement of a new
fundamental physical constant. It is deplorable, that LEP-2 was not run at a larger
energy and a longer period. Clarity about this model can only come
with a new lepton machine, the one most studied being the linear collider,
which however might not be an optimal machine for this type of model,
because of the beamstrahlung. However more detailed studies are needed here. 
From the theoretical point of view it is interesting, that the higher dimensional
field could be massless without spoiling renormalizability, which would actually
imply the absence of any particle pole.

As this type of models is rather unfamiliar one might be somewhat
sceptical about their meaning. A typical objection is that we have
never seen a fundamental field with a non-trivial \KL\, spectrum,
so it would be unlikely that the Higgs-boson would be different. However
one can easily make the objection, that we also have never seen a 
fundamental scalar field, so why should the Higgs field behave
as other particles. The special place of the Higgs field in the theory
is an argument in point of this view. Also the $\sigma$-resonance
in the strong interactions looks quite similar to the Higgs field here.

One can of course take a more positive point of view. After all the original Higgs field
was introduced to solve a fundamental problem. We know massive gauge
vectorbosons exist, but by themselves they cannot have a unitary
scattering matrix. In order to cure this problem a field with the same 
quantum numbers as the Higgs is needed. It cannot be too heavy, since that
would violate unitarity. Since we assume relativistic quantum field theory to be correct,
such a field must carry a \KL\, spectral density. But nothing tells us, that it has
to be a single particle pole. Generally speaking that would not  be the case.
So assuming the existence of a Higgs boson field with an unknown spectral density,
is the minimal assumption that one must make, in order to cure the problem
of unitarity. Everything else is a stronger assumption, whether well or not so well
motivated.
Actually what this paper shows, is that even in the restricted class of
renormalizable models with a single standard model Higgs, there is a broad
choice of possibilities for a non-trivial spectrum.
One should therefore accept this type of model as the generic form of the
simplest Higgs mechanism, the single particle being a  special case.
This would also mean, that the LHC is not the right machine to study
the mechanism of electroweak symmetry breaking. However it is a very
good machine to rule out exotic models; it is potentially an almost perfect 
 null-experiment for such a purpose.
If the LHC finds no signal for new physics, not even for the Higgs particle, 
this would basically prove that the HEIDI models are the right class of models to
describe electroweak symmetry breaking.
This would provide an exceedingly strong physics argument for building a
lepton collider. Such a collider  would not need a very high energy; 300\gev\, would be
more than enough.
One needs however a high luminosity and a high precision on the beam energy.
It is somewhat ironic, that this type of model was not discussed already
in the 1960's, when the Higgs particle was introduced, as in this period
much research was done on dispersion relations, which is a subject
closely related to the \KL\, spectral density.
Besides in the Higgs sector the idea of mixing a four-dimensional field with a 
high dimensional singlet in a renormalizable way works also for (abelian) vector-fields
\cite{lorca,fuks,krasnikovz} 
 or for right-handed neutrinos. Necessary is that the 
four-dimensional fields
are singlets under the gauge group; otherwise renormalizability is lost. \\

{\bf Acknowledgements} We thank Dr. O.~Brein for discussions and a careful
reading of the manuscript. This work was supported by the DFG within
the Graduiertenkolleg "Physik an Hadron-Beschleunigern".


\begin{thebibliography}{r}

\bibitem{lephiggs}ALEPH, DELPHI, L3 and OPAL Collaborations
and the LEP Working Group for Higgs Boson Searches,
 Phys. Lett. B {\bf 565}, 61 (2003). 
\bibitem{kane}
  G.~L.~Kane, T.~T.~Wang, B.~D.~Nelson and L.~T.~Wang,
  %``Theoretical implications of the LEP Higgs search,''
  Phys.\ Rev.\  D {\bf 71}, 035006 (2005).
%  [arXiv:hep-ph/0407001];
  %%CITATION = PHRVA,D71,035006;%%

\bibitem{drees}
  M.~Drees,
  %``A Supersymmetric explanation of the excess of Higgs-like events at LEP,''
  Phys.\ Rev.\  D {\bf 71}, 115006 (2005).
%  [arXiv:hep-ph/0502075];
  %%CITATION = PHRVA,D71,115006;%%

\bibitem{hooper}
  D.~Hooper and T.~Plehn,
  %``Dark matter and collider phenomenology with two light supersymmetric  Higgs
  %bosons,''
  Phys.\ Rev.\  D {\bf 72}, 115005 (2005).
% [arXiv:hep-ph/0506061];
  %%CITATION = PHRVA,D72,115005;%%

\bibitem{demir}
  D.~A.~Demir, L.~Solmaz and S.~Solmaz,
  %``LEP indications for two light Higgs bosons and U(1) -prime model,''
  Phys.\ Rev.\  D {\bf 73}, 016001 (2006).
%  [arXiv:hep-ph/0512134].
  %%CITATION = PHRVA,D73,016001;%%

\bibitem{hill}
  A.~Hill and J.~J.~van der Bij,
  %``STRONGLY INTERACTING SINGLET - DOUBLET HIGGS MODEL,''
  Phys.\ Rev.\  D {\bf 36}, 3463 (1987).
  %%CITATION = PHRVA,D36,3463;%%
\bibitem{mnmsm}J.J. van der Bij, Phys.\ Lett.\ B {\bf 636}, 56 (2006).

\bibitem{dilchbij}
  J.~J.~van der Bij and S.~Dilcher,
  %``A higher dimensional explanation of the excess of Higgs-like events at CERN
  %LEP,''
  Phys.\ Lett.\  B {\bf 638}, 234 (2006).
%  [arXiv:hep-ph/0605008].
  %%CITATION = PHLTA,B638,234;%%


\bibitem{kallen}
G.~K\"all\'en, Helv. Phys. Acta  {\bf 25}, 417 (1952).
\bibitem{lehmann}
H.~Lehmann, Nuovo Cimento {\bf 11}, 342 (1954). 
\bibitem{georgi} H. Georgi, Phys.\ Rev.\ Lett. {\bf 98}, 221601 (2007).
\bibitem{heidiun} J.~J.~van~der~Bij and S.~Dilcher, Phys.\ Lett.\ B {\bf 655}, 183 (2007).

\bibitem{wise} W.D. Goldberger, M.B. Wise,
 Phys. Rev. D {\bf 65}, 025011 (2001).
\bibitem{dilcher} S. Dilcher, Doktorarbeit Freiburg (2001).
\bibitem{aguila} F. del Aguila, M. P\'erez-Victoria, J. Santiago,
         JHEP0302, 051 (2003).
\bibitem{derham} C.~de~Rham, JHEP0801, 060 (2008).

\bibitem{lorca} A. Ferroglia, A. Lorca and J.J. van der Bij,
      Ann. Phys. (Leipzig) {\bf 16}, 563 (2007).
%      hep-ph/0611174 (2006).
\bibitem{fuks}B.~Fuks, Qingjun~Xu and J.~J.~van~der~Bij,
      Phys. Rev. D {\bf 78}, 074016 (2008).
\bibitem{krasnikovz} N.~V.~Krasnikov, Mod. Phys. Lett. {\bf A25}, 2313 (2010).




\end{thebibliography}
\end{document}